\documentclass[%
onecolumn,%
oneside,%
floats,%
aps,%
prd,%
nobibnotes,%
nofootinbib,%
amsmath,%
amssymb,%
amsfonts,%
amscd,%
showpacs,%
superscriptaddress,%
eqsecnum%
]{revtex4}

\usepackage[utf8]{inputenc}
\usepackage{graphicx,array,dcolumn}
\usepackage[paperwidth=210mm,paperheight=297mm,centering,hmargin=1.8cm,vmargin=2.5cm]{geometry}

\newcommand{\be}{\begin{equation}}
\newcommand{\ee}{\end{equation}}

\def\b{\textbf}
\def\i{\textit}
\def\mb{\mathbf}
\def\mc{\mathcal}
\def\be{\begin{equation}}
\def\ee{\end{equation}}
\def\bea{\begin{eqnarray}}
\def\eea{\end{eqnarray}}
\def\mpl{m_{Pl}}
\def\sun{\odot}
\def\mn{{\mu\nu}}
\def\D{\mc D}
\def\-g{\sqrt{-g}}

\renewcommand\rho{\varrho}
\renewcommand\tilde{\widetilde}
\newcommand\rif[1]{(\ref{#1})}
\def\so{\Rightarrow}

\begin{document}


\title{Particle Production in $f(R)$ Gravity during Structure Formation}

\author{E.V. Arbuzova}
\email{arbuzova@uni-dubna.ru}
\affiliation{Novosibirsk State University, Novosibirsk, 630090, Russia}
\affiliation{Department of Higher Mathematics, University "Dubna", 141980 Dubna, Russia}

\author{A.D. Dolgov}
\email{dolgov@fe.infn.it}
\affiliation{Novosibirsk State University, Novosibirsk, 630090, Russia}
\affiliation{ITEP, Bol. Cheremushkinsaya ul., 25, 113259 Moscow, Russia}
\affiliation{Dipartimento di Fisica e Scienze della Terra, Universit\`a degli Studi di Ferrara\\
Polo Scientifico e Tecnologico - Edificio C, Via Saragat 1, 44122 Ferrara, Italy}
\affiliation{Istituto Nazionale di Fisica Nucleare (INFN), Sezione di Ferrara\\
Polo Scientifico e Tecnologico - Edificio C, Via Saragat 1, 44122 Ferrara, Italy}

\author{L. Reverberi}
\email{reverberi@fe.infn.it}
\affiliation{Dipartimento di Fisica e Scienze della Terra, Universit\`a degli Studi di Ferrara\\
Polo Scientifico e Tecnologico - Edificio C, Via Saragat 1, 44122 Ferrara, Italy}
\affiliation{Istituto Nazionale di Fisica Nucleare (INFN), Sezione di Ferrara\\
Polo Scientifico e Tecnologico - Edificio C, Via Saragat 1, 44122 Ferrara, Italy}


\begin{abstract}
We study particle production in infrared-modified gravitational theories in the
contemporary universe. It is shown that in astronomical systems with rising mass density, the 
curvature scalar may oscillate with very high frequency.
These oscillations lead to efficient particle production and in an interesting range of model 
parameters could be a source of energetic cosmic rays. This effect either excludes 
some models of modified gravity or suggests a new mechanism of cosmic ray production.
\end{abstract}

\maketitle

\section{Introduction}

The discovery of the present-day accelerated cosmological expansion \cite{Nobel_2011} has triggered huge theoretical efforts directed to finding an explanation for its value and for its very existence, but the driving force behind this phenomenon is still unknown. A popular possibility is represented by gravity modifications at large scales and/or small curvatures, which can be realized by adding a non-linear function of the scalar curvature $R$ to the usual Einstein-Hilbert action:\footnote{We adopt the conventions $c=\hbar=k=1$, $R_\mn = R^\alpha_{\,\,\mu\alpha\nu}$, $R^\alpha_{\,\,\mu\beta\nu}=\partial_\beta\Gamma^\alpha_\mn-\cdots$; the metric signature is $(+\,-\,-\,-)$.}
\be
A_{grav}=-\frac{\mpl^2}{16\pi}\int d^4x\-g\,f(R)\equiv -\frac{\mpl^2}{16\pi}\int d^4x\-g\,\left[R+F(R)\right]\,.
\ee
The initial suggestion~\cite{one-over-R} of gravity modification with $F(R) \sim \mu^4/R$
suffered from strong instabilities in celestial bodies~\cite{DolgKaw}. Because of that, further modifications have been 
suggested~\cite{HuSaw,ApplBatt,Starobinsky_2007} which are free of these instabilities. For a review of these and some other versions 
see e.g.~\cite{rev-f-of-R,noj-odin}. The suggested modifications, however, may lead to infinite-$R$ singularities 
in the past cosmological history~\cite{rev-f-of-R} and in the future in astronomical systems
with rising energy/matter density~\cite{frolov, Arb_Dolgov}. Some properties of the singularity found  
in \cite{Arb_Dolgov} were further studied in \cite{bamba-noj-odin}.
These singularities can be successfully cured
by the addition of an $R^2$-term into the action. Such a contribution naturally appears as a result of  
quantum corrections due to matter loops in curved space-time~\cite{Gur-Star,Starobinsky_1980}.
Another mechanism which may in principle eliminate these singularities is particle production by the
oscillating curvature. If the production rate is sufficiently high, the oscillations of $R$ are efficiently damped and the singularity could be avoided (see below).

The $R^2$ term may also have dominated in the early universe where it could lead to strong particle
production. The process was studied long ago in~\cite{Zeld-Star, Starobinsky_1980,Vilenkin_1985}.
Renewed interest to this problem arose recently~\cite{Arb_Dolg_Rev,japanese}, stimulated by the interest in
possible effects of additional ultraviolet terms, $\sim R^2$, in infrared-modified $F(R)$ gravity models. 
A general approach to the origin of singularities in $F(R)$ theories was considered in ref.~[18].

In this paper we discuss the behaviour of a popular $F(R)$ model of dark energy in the case of a contracting system, discussing the 
evolution of the curvature scalar $R$ and the related effects of gravitational particle production.
The calculations are done both numerically and analytically. For realistic values of the parameters, especially for extremely small
coupling constant $g$, see Eq.~ (\ref{eq:definitions}), numerical calculations are not reliable, so we have found an approximate
 analytical solution and compared it with numerical one with small but not too small values of $g$, for which numerical solutions are reliable. The comparison confirms the very good precision of the analytical solution.

\section{Basic Frameworks and Equations \label{s-basics}}

We consider the model proposed in ref.~\cite{Starobinsky_2007}:
\be\label{eq:model}
F(R) = -\lambda R_c\left[1-\left(1+\frac{R^2}{R_c^2}\right)^{-n}\right]-\frac{R^2}{6m^2}\,,
\ee
where $n$ is an integer, $\lambda>0$, and $| R_c |$ is of the order of $8\pi \rho_c / m_{Pl}^2$, where
$\rho_c$ is the present day value of the total cosmological energy density. More precisely the
value of $R_c$ is determined by equation (\ref{eq-Rc}) below.
The $R^2$-term, absent in the original formulation, has been included to prevent curvature singularities in the presence of 
contracting bodies \cite{Arb_Dolgov}, and is relevant only at very large curvatures, because we need $m\gtrsim 10^5$ GeV in order to preserve the successful predictions of the standard BBN \cite{Arb_Dolg_Rev}.

Though we made explicit calculations for the model of ref.~\cite{Starobinsky_2007}, the results are applicable
to similar modifications proposed in papers~\cite{HuSaw,ApplBatt} with the addition of the $R^2$ term, though 
precise numerical values may be somewhat different. The point is that the evolution of $R$ in all three suggested forms of  $F(R)$ is determined by the behavior of the potential $U$ defined below in eqs.~\rif{U-prime} or~\rif{eq:xi_potential} and this behavior is qualitatively the same.

The evolution of $R$ is determined from the trace of the modified Einstein equations:
\be\label{eq:trace}
3\D^2 F_{,R} -R+RF_{,R} -2F=T\,,
\ee
where $\D^2\equiv \D_\mu\D^\mu$ is the covariant D'Alambertian operator, $F_{,R} \equiv d F/ d R$,  $T~\equiv~8\pi T^\mu_\mu/\mpl^2$, and $T_{\mu\nu}$ is the energy-momentum tensor of matter.

To describe the accelerated cosmological expansion, the function $F(R)$ is chosen in such a way that equation (\ref{eq:trace}) has a non-zero
constant curvature solution, $R=\bar R $, in the absence of matter. Observational data demand
\be
\bar R = - \frac{32\pi \Omega_\lambda \rho_c}{m_{Pl}^2},
\label{R-c}
\ee
where $\Omega_\lambda \approx 0.75$ is the vacuum-like cosmological energy density, deduced from the observations under the assumption of validity of the usual General Relativity (GR) with non-zero cosmological constant. Using this condition we can determine
$R_c$ from the solution of the equation:
\be
\bar R-\bar R F_{,R}  (\bar R) + 2 F(\bar R) = 0.
\label{eq-Rc}
\ee
This equation has two different limiting solutions for sufficiently large $\lambda$, roughly speaking $\lambda > 1$, namely $\bar R/R_c = 2\lambda$ and
$\bar R/R_c =1/ [n(n+1)\lambda]^{1/3}$. Following ref.~\cite{Starobinsky_2007}, we should consider only the \i{maximal} root $\bar R\lesssim 2\lambda R_c$. Moreover, for the sake of simplicity and definiteness, we will neglect these subtleties and assume $\lambda \sim 1$ and
\be
R_c \simeq \bar R \simeq 1/t_U^2,
\label{R-0}
\ee
where $t_U \approx 4\cdot 10^{17}$ s is the universe age. Still, for a more detailed study of the parameter space of the model, it could be necessary to consider the full numerical solution of Eq.~\rif{eq-Rc} for all values of $\lambda$.

We are particularly interested in the regime $|R_c|\ll|R|\ll m^2$, in which $F$ can be approximated by
\be\label{eq:model_approx}
F(R)\simeq -R_c\left[1-\left(\frac{R_c}{R}\right)^{2n}\right]-\frac{R^2}{6m^2}\,.
\ee
We consider a nearly-homogeneous distribution of pressureless matter, with energy/mass density rising with time but still relatively low (e.g. a gas cloud in the process of galaxy or star formation). In such a case the spatial derivatives can be neglected and, if the object is far from forming a black hole, 
the space-time would be almost Minkowski. Then equation \rif{eq:trace} takes the form
\be\label{eq:trace_approx}
3\partial_t^2F_{,R} -R-T = 0\,.
\ee
Let us introduce the dimensionless quantities\footnote{The parameter $g$ should not be confused with $\det g_\mn$.}
\be\label{eq:definitions}
\begin{gathered}
 z\equiv \frac{T(t)}{T(t_{in})}\equiv \frac{T}{T_0}= \frac{\rho_m(t)}{\rho_{m0}}\,,
 \qquad y\equiv -\frac{R}{T_0}\,, \\
g\equiv \frac{T_0^{2n+2}}{6 n(-R_c)^{2n+1}m^2}= \frac{1}{6 n  (m t_U)^2} \,\left( \frac{\rho_{m0}}{\rho_c}\right)^{2n+2}\,,
\qquad \tau\equiv m\sqrt g\,t\,,
\end{gathered}
\ee
where $\rho_c \approx 10^{-29} $ g/cm$^3$ is the cosmological energy density at the present time,
$\rho_{m0}$ is the initial value of the mass/energy density of the object under scrutiny,
and $T_0 = 8\pi \rho_{m0}/m_{Pl}^2$. Next let us introduce the new scalar field:
\be\label{eq:xi_definition}
\xi\equiv  \frac{1}{2 n}\left(\frac{T_0}{R_c}\right)^{2n+1}F_{,R}  = \frac{1}{y^{2n+1}}-gy\,,,
\ee
in terms of which Eq.~\rif{eq:trace_approx} can be rewritten in the simple oscillator form:
\be\label{eq:xi_evol}
\xi''+z-y=0\,,
\ee
where a prime denotes derivative with respect to $\tau$. The potential of the oscillator is defined by:
\be
\frac{\partial U}{\partial \xi}= z - y(\xi).
\label{U-prime}
\ee
The substitution (\ref{eq:xi_definition}) is analogous to that done in \cite{Arb_Dolgov} but now $y$ cannot be 
analytically expressed through $\xi$ and we have to use approximate expressions.

It is clear that~\rif{eq:xi_evol} describes oscillations around $y=z$ (the ``bottom'' of the potential), 
which corresponds to the usual GR solution $R+T=0$. So we can separate solutions into an average and an oscillatory part. 
For small deviations from the minimum of the potential, solutions take the form:
\be
\xi(\tau)=\left[\frac{1}{z(\tau)^{2n+1}}-gz(\tau)\right]+\alpha(\tau)\sin F(\tau)\equiv \xi_a(\tau)+\xi_1(\tau)\,,
\label{eq:xi_expansion}
\ee
where
\be
F (\tau) \equiv \int^\tau_{\tau_0}d\tau'\,\Omega (\tau')\,,
\ee
and the dimensionless frequency $\Omega$ is defined as
\be\label{eq:frequency_U}
\Omega^2 = \frac{\partial^2 U}{\partial \xi^2}\,,
\ee
taken at $y=z$. From \rif{eq:xi_evol}, we find that it is equal to
\be\label{eq:frequency_2}
\Omega^2 = -\left.\frac{\partial y}{\partial\xi}\right|_{y=z} = -\left.\frac{1}{\partial\xi/\partial y}\right|_{y=z} = \left(\frac{2n+1}{z^{2n+2}}+g\right)^{-1}\,.
\ee
The conversion into the physical frequency $\omega$ is given by
\be\label{eq:Omega}
\omega = \Omega\,m \sqrt g\,.
\ee
It is assumed that initially $\xi (\tau_0)$ sits at the minimum of the potential, otherwise we would need to add a cosine term in~\rif{eq:xi_definition}. 
If initially $\xi (\tau_0)$ was shifted from the minimum, the oscillations would generally be stronger and the effect of particle 
production would be more pronounced.

\subsection{Potential for $\xi$}
\begin{figure}
\centering
 \includegraphics[width=.4\textwidth]{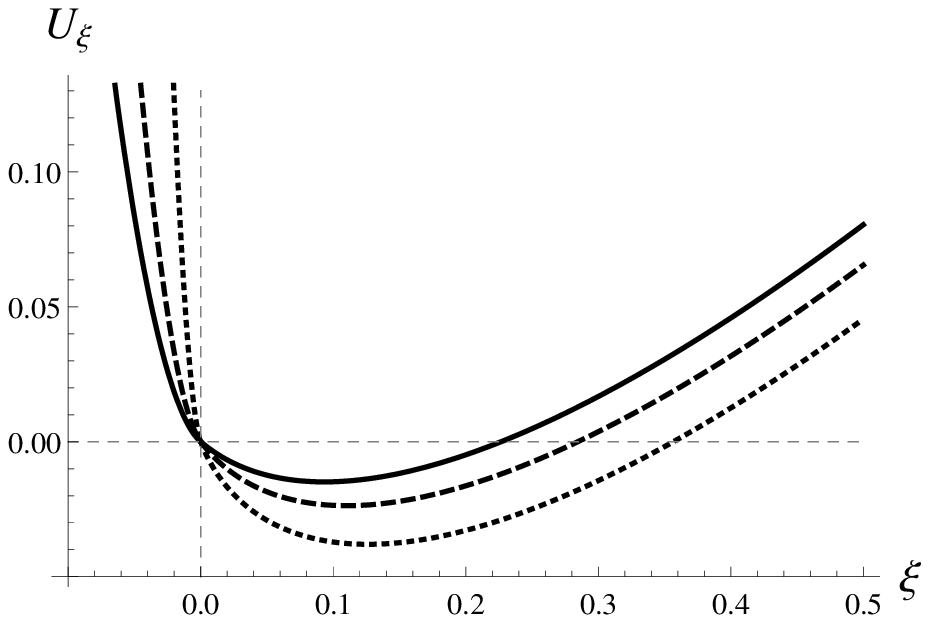}
\includegraphics[width=.4\textwidth]{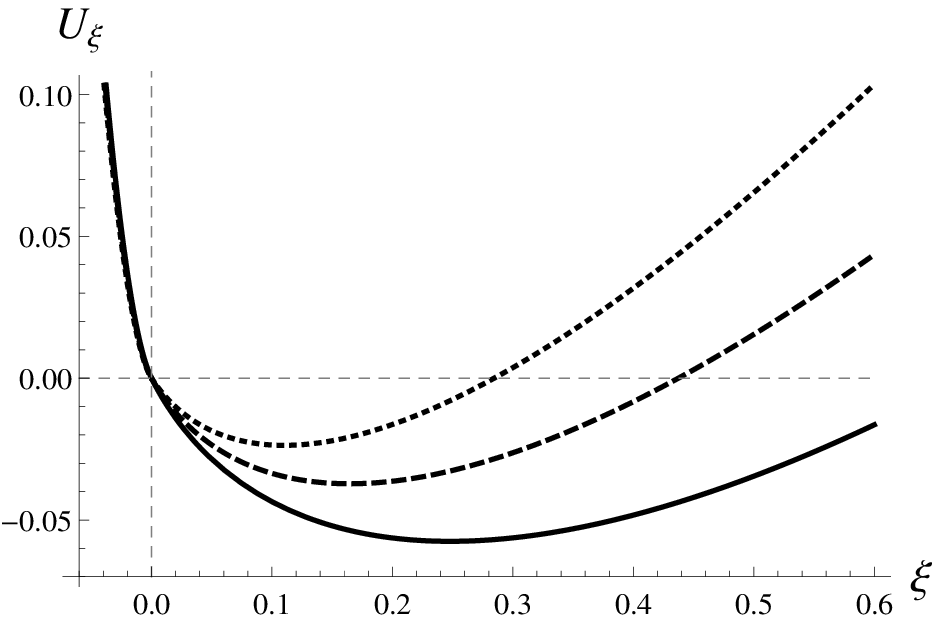}
\caption{Examples of the variation of potential~\rif{eq:xi_potential} for different values of parameters. 
\i{Left panel} $(n=2,\,z=1.5)$ solid line: $g=0.02$, dashed line: $g=0.01$, dotted line: $g=0.002$; the part of the potential at $\xi<\xi_a$ 
is increasingly steeper as $g$ decreases; the bottom of the potential also moves. \i{Right panel} $(n=2,\,g=0.01)$ solid line: $z=1.3$, 
dashed line: $z=1.4$, dotted line: $z=1.5$; the bottom of the potential moves to higher values of $U$ and lower values of $\xi$, 
as $z$ increases.}
\label{fig:potential}
\end{figure}

One cannot analytically invert Eq.~\rif{eq:xi_definition} to find the exact expression for $U(\xi)$. However, we can find an approximate expression for $gy^{2n+2}\ll 1$ ($\xi>0$) and $gy^{2n+2}\gg 1$ ($\xi<0$). The value $\xi=0$ separates two very distinct regimes, in each of which $\Omega$ 
has a very simple expression [see Eq.~\rif{eq:frequency_2}] and $\xi$ is dominated by either one of the two terms in the r.h.s. 
of Eq. \rif{eq:xi_definition}. Hence, in those limits the relation $\xi=\xi(y)$ can be inverted giving an explicit expression for $y=y(\xi)$, and therefore the following form for the potential:
\begin{subequations}\label{eq:xi_potential}
\be\label{eq:xi_pot_theta}
U(\xi) = U_+(\xi)\Theta(\xi) + U_-(\xi)\Theta(-\xi)\,,
\ee
where
\be\label{eq:potential_+_-}
\begin{aligned}
U_+(\xi) &= z\xi - \frac{2n+1}{2n}\left[\left(\xi+g^{(2n+1)/(2n+2)}\right)^{2n/(2n+1)}-g^{2n/(2n+2)}\right]\,,\\
U_-(\xi) &= \left(z-g^{-1/(2n+2)}\right)\xi+\frac{\xi^2}{2g}\,.
\end{aligned}
\ee
\end{subequations}
By construction $U$ and $\partial U/\partial\xi$ are continuous at $\xi=0$. The shape of this potential is shown in Fig.~\ref{fig:potential}.
We can write a conservation equation for a quantity which is analogous to the ``energy'' of the field $\xi$:
\be\label{eq:conserved}
\frac{1}{2}\,\xi'^2 + U(\xi) - \int^\tau_{\tau_0}d\eta \frac{\partial U}{\partial\eta} = \frac{1}{2}\,\xi'^2 + 
U(\xi) - \int^\tau_{\tau_0}d\eta \frac{\partial z}{\partial\eta}\,\xi(\eta)=\text{ const}\,,
\ee
where $\xi$ and $\xi'$ are taken at time moment $\tau$ coinciding with the upper integration bound.
The oscillating part of $\xi$ in the last integral term in~\rif{eq:conserved} would be integrated away for  fast harmonic oscillations of $\xi$. However, since the oscillations 
at late time become  strongly asymmetric, this term rises with time, see Fig. 3 and Sec. 3.3.1 below.

The bottom of the potential, as it is obvious from Eq.~\rif{U-prime}, corresponds to the GR solution $R=-T$, or $y(\xi) = z$, and its depth (for $gz^{2n+2}<1$) is
\be\label{eq:potential_bottom}
U_0(\tau) \simeq -\frac{1}{2n\,z(\tau)^{2n}}\,.
\ee

We will use a very simple form for the external energy density $z$, namely
\be\label{eq:z_linear}
\begin{aligned}
z(\tau) &= 1+\kappa(\tau-\tau_0)\\
\rho(t)&= \rho_0\left(1+\frac{t-t_0}{t_{contr}}\right)\\
\kappa^{-1} &\equiv m\sqrt g\,t_{contr}\,.
\end{aligned}
\ee
Here, $\kappa^{-1}$ and $t_{contr}$ are respectively the dimensionless and physical timescales of the contraction of the system; analogously, $\tau_0$ and $t_0$ are respectively the dimensionless and physical initial times, which for simplicity and without loss of generality will be taken equal to 0.
This evolution law may not be accurate when $t/t_{contr}>1$, but results obtained with more sophisticated functions describing the contraction of the system are most likely in qualitative agreement with our results, provided that $\dot\rho$ remains positive at all times.

It is also useful to express physical parameters such as $m$,  the initial energy density $\rho_{m0}$, etc., in terms of their respective ``typical'' values. Let us define
\be\label{eq:typ_param}
\begin{aligned}
 &\rho_{29}\equiv \frac{\rho_{m0}}{\rho_c}\,,\\
&m_5 \equiv \frac{m}{10^5\text{ GeV}}\,,\\
&t_{10}\equiv \frac{t_{contr}}{10^{10}\text{ years}}\,,
\end{aligned}
\ee
where $\rho_c=10^{-29}\text{ g cm}^{-3}$ is the present (critical) energy density of the Universe. In terms of these quantities, we can rewrite $g$ and $\kappa$ as
\be
\begin{aligned}
g &\simeq 1.2\times 10^{-94}\,\frac{\rho_{29}^{2n+2}}{n 
\,m_5^2}\,,\\
\kappa &\simeq 1.9 \,\frac{\sqrt{n}
}{\rho_{29}^{n+1}\,t_{10}}\,.
\label{g-kappa}
\end{aligned}
\ee

\section{Solutions}

\subsection{Oscillations of $\xi$}
\begin{figure}
\centering
\includegraphics[width=.4\textwidth]{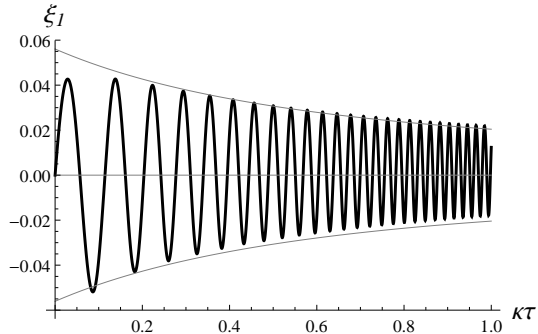}
\caption{Numerically calculated oscillations of $\xi_1(\tau)$, in the case $n=2$, $\kappa=0.01$, $g=0.01$ and initial conditions $y_0=1$, 
$y'_0=\kappa/2$. 
The amplitude and the frequency of the oscillations are in very good agreement with our analytical result~\rif{eq:xi_amplitude_solution}.}
\label{fig:xi_amplitude}
\end{figure}

At first order in $\xi_1$, equation \rif{eq:xi_evol} can be written  as
\be\label{eq:xi_evol_approx}
\xi_1''+\Omega^2\xi_1= -\xi_a'' \,,
\ee
with $\Omega$ given by~\rif{eq:frequency_2}. The term $\xi_a''$ is proportional to $\kappa^2$, which is usually 
assumed small, so in first approximation it can be neglected, though an analytic solution for constant $\Omega$ or in the limit of large $\Omega$ can be obtained considering this term as well. Using~\rif{eq:xi_expansion} and neglecting $\alpha''$, we obtain
\be\label{eq:alpha_omega}
2\,\frac{\alpha'}{\alpha} \simeq -\frac{\Omega'}{\Omega}\so \alpha\simeq \alpha_0\,
\sqrt\frac{\Omega_0}{\Omega}= \alpha_0\left(\frac{1}{z^{2n+2}}+\frac{g}{2n+1}\right)^{1/4}
\left({1}+\frac{g}{2n+1}\right)^{-1/4}.
\ee
Here and in what follows sub-0 means that the corresponding quantity is taken at initial moment $\tau = \tau_0$. We impose the following initial conditions
\be
\begin{cases}
y(\tau=\tau_0)=z(\tau=\tau_0)=1\,,\\
y'(\tau=\tau_0)=y'_0\,,
\end{cases}
\label{init-y}
\ee
which correspond to the GR solution at the initial moment. In terms of $\xi$  it means that ${\xi_1(\tau_0)=0}$. 
The initial value of the derivative $\xi'_1 (\tau_0)$ can be expressed through $y_0'$, which we keep as a free parameter; 
according to~\rif{eq:xi_definition}: $\xi'_0 = -y'_0 (2n+1+g) $. Differentiating Eq.~(\ref{eq:xi_expansion}) with respect to $\tau$ and using \rif{eq:alpha_omega} we find:
\be 
\alpha_0 = (\kappa - y'_0) (2n+1+g)^{3/2}.
\label{alpha-0}
\ee
Correspondingly:
\be\label{eq:xi_amplitude_solution}
\left|\alpha(\tau)\right| = |y'_0-\kappa|(2n+1+g)^{5/4}\left(\frac{2n+1}{z^{2n+2}}+g\right)^{1/4}\,.
\ee
Because of the assumptions made to obtain \rif{eq:alpha_omega}, we expect this result to hold when $|y'_0-\kappa|\sim\kappa$ or slightly less. 
In this regime the numerical results, shown in Fig.~\ref{fig:xi_amplitude}, are in excellent agreement with the analytical 
estimate~\rif{eq:xi_amplitude_solution}. 
We remark that the agreement improves for larger $g$  and/or smaller $\kappa$, while for small $g$ and ``large'' $\kappa$ it may become 
significantly worse (see paragraph \ref{sec:spikes}).

For $ y'_0 = \kappa$ and $ \xi_1 (\tau_0)=0$, it would seem from Eq.~\rif{eq:xi_amplitude_solution} that oscillations are not excited. However, this is an artifact of the approximation used. In fact, the "source" term in the r.h.s. of Eq.~\rif{eq:xi_evol_approx} produces oscillations and hence deviations from GR
with any initial conditions.

\subsection{Oscillations of $y$}
We shall now exploit this result to evaluate the amplitude of the oscillations of $y$. We first expand $y$ as it was done for $\xi$ in Eq.~\rif{eq:xi_expansion}:
\be\label{eq:y_expansion}
y(\tau) = z(\tau) + \beta(\tau)\,\sin F(\tau) \equiv y_a(\tau) + \beta(\tau)\,\sin\left(\int_{\tau_0}^\tau d\tau'\,\Omega \right)\,, 
\ee
where it is easy to prove that $\Omega$ must coincide with that given by Eq.~\rif{eq:frequency_2}. For $|\beta/z| < 1$, we expand $\xi$ as
\begin{align}\label{eq:xi_exp_small_beta}
\xi&=\frac{1}{z^{2n+1}\left[1+({\beta}/{z})\sin F(\tau)\right]^{2n+1}}-g\left(z+\beta\sin F(\tau)\right)\notag\\
 &\simeq \frac{1}{z^{2n+1}}-gz - \left(\frac{2n+1}{z^{2n+2}}+g\right)\beta\sin F(\tau)
\end{align}
Comparing this expression with Eqs.~\rif{eq:xi_expansion} and \rif{eq:frequency_2}, we find that
\be
|\beta| = |\alpha|\left(\frac{2n+1}{z^{2n+2}}+g\right)^{-1} = |\alpha|\Omega^2\,.
\ee
Accordingly, $\beta$ evolves as:
\be\label{eq:y_amplitude_evolution}
\beta(\tau) \simeq \left|y'_0-\kappa\right|\left(2n+1+g\right)^{5/4}\left(\frac{2n+1}{z^{2n+2}}+g\right)^{-3/4}\,.
\ee
This is in reasonable agreement with numerical results, especially in both limiting cases ${gz^{2n+2}\ll 1}$ and ${gz^{2n+2}\gg 1}$, as expected.

\subsection{``Spike-like'' Solutions}\label{sec:spikes}
We have found simple analytical solutions for $\xi$ and $y$ in two separate limits: $gz^{2n+2}\ll 1$ and $gz^{2n+2}\gg 1$. 
However, in the intermediate case numerical calculations show interesting features which are worth discussing.
\begin{figure}[!t]
\centering
 \includegraphics[width=.4\textwidth]{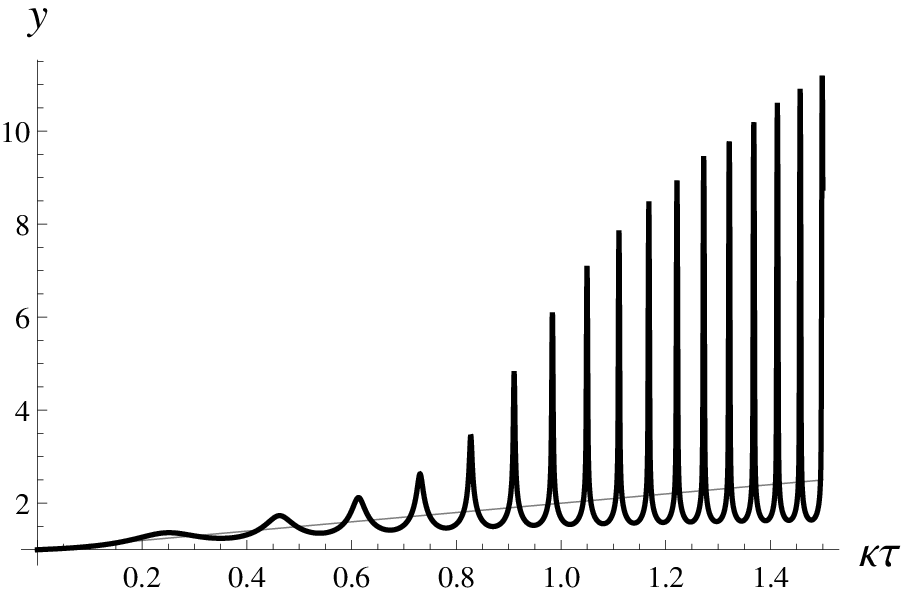}
\includegraphics[width=.4\textwidth]{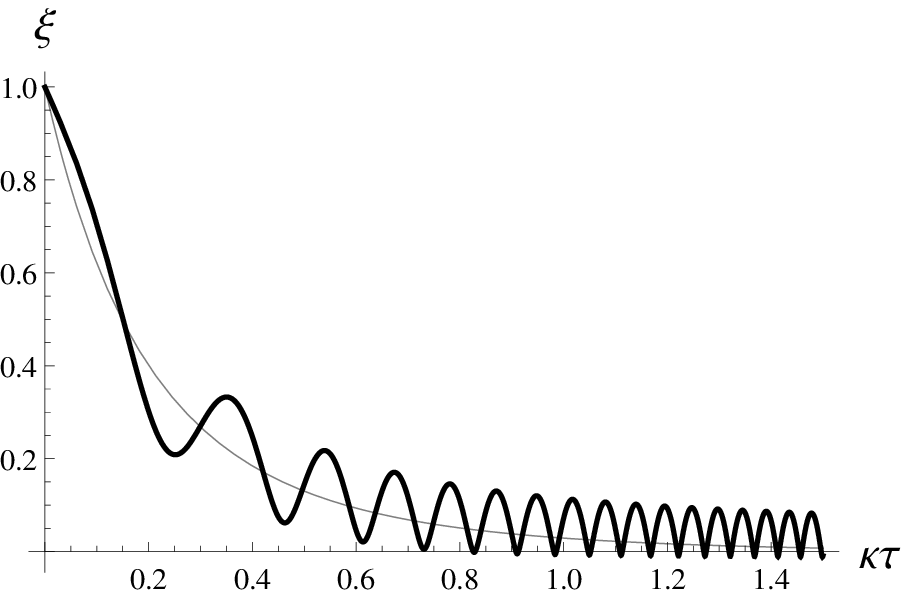}
\caption{``Spikes'' in the solutions. The results presented are for $n=2$, $g=0.001$, $\kappa=0.04$, and $y'_0=\kappa/2$. 
Note the asymmetry of the oscillations of $y$ and $\xi$ around $y=z$ and their anharmonicity. See the text for further details.}
\label{fig:spikes}
\end{figure}
As shown in figure \ref{fig:spikes}, when $\xi$ approaches (and even crosses) zero but $\xi_a$ does not, that is when $gy^{2n+2}\simeq 1$ but $gz^{2n+2}<1$, we have the largest deviations from the harmonic, symmetric oscillations around $y=z$. This happens especially when $g$ is very small and $\kappa$ is not too small.\\
The reason for this behaviour is qualitatively explained by the following considerations. Inspecting  Eq.~\rif{eq:xi_potential} and/or figure \ref{fig:potential}, we see that when $\xi<\xi_a$ the potential becomes increasingly steep, hence reducing the time spent in that region. Moreover, a given variation $\delta\xi$ in this region corresponds to a large variation of $y$. Thus there appear high, narrow ``spikes'' in $y$. On the other hand, for $\xi>\xi_a$ the potential is much less steep, and the oscillation in that region lasts longer, yielding slow ``valleys'' between the spikes of $y$.\\
Please note that in the region with spikes, the assumption $|\beta/z| \ll 1$ is no longer accurate, and we have deviations from the analytical estimate~\rif{eq:y_amplitude_evolution}, which is usually smaller than the exact numerical value.

\subsubsection{Estimate of $\beta$}\label{sec:beta_estimate}

In order to obtain an estimate of $\beta$ in this region we use an analogous approach to that of~\cite{Reverberi2013}. Let us first introduce some new notations:
\be 
U(\tau) = U[\xi(\tau), z(\tau)], \,\,\,\, U_a(\tau) = U[\xi_a (\tau), z(\tau)].
\label{z-of-tau}
\ee
Remember that according to Eq.~(\ref{eq:xi_expansion}) $\xi_a (\tau)  =  z(\tau)^{-(2n+1)} - g z(\tau)$ is the value of $\xi$ where the potential has its minimum value. Let us denote by $\tau_a$ the time at which $\xi$ hits  the minimum of the potential, i.e. 
$\xi (\tau_a) = \xi_a (\tau_a)$ and correspondingly $\xi_1 (\tau_a) = 0$. Note that $U(\tau_a) =U_a(\tau_a)$, while they are evidently different
at other values of $\tau$. We will examine Eq.~\rif{eq:conserved} choosing the initial value of time at the moment when $\xi$ passes through the
minimum of the potential. Correspondingly $U(\tau_0) = U_a (\tau_0)$ and hence we denote $\tau_0$ as $ \tau_{a0}$. 
With this initial value of $\tau_0$ we rewrite Eq.~(\ref{eq:conserved}) as
\be\label{new_conserved}
\frac{1}{2} [\xi'(\tau)]^2+U(\tau)-\int^\tau_{\tau_{a0}} d\eta\,z'(\eta) \left[ \xi_a(\eta) + \xi_1(\eta)  \right]= U_a(\tau_{a0}) + \frac{1}{2} [\xi'(\tau_{a0})]^2\,.
\ee
We start by noticing that the first term under the integral in the l.h.s. of this equation can be explicitly integrated and is equal to 
\be\label{eq:int_source_potential}
\int^\tau_{\tau_{a0}} d\eta\,z' (\eta) \,\xi_a (\eta)= U_a(\tau) - U_a(\tau_{a0})\,.
\ee
So $U_a(\tau_{a0})$ disappears from this equation.

Assume that the upper limit of integration, $\tau$, is sufficiently high, such that the nearest minimum value of $\xi$ is negative.
Let us first take $\tau$ exactly equal to $\tau_m$, i.e. to one of the numerous  values when $\xi$ hits a minimum, $\xi = \xi_{min}<0$.
At this point $\xi'(\tau_m) = 0$ and we obtain:
\be
U(\tau_m) = U_a(\tau_m) + \frac{1}{2}\, [\xi'(\tau_{a0})]^2 + \int^{\tau_m}_{\tau_{a0}} d\eta z'(\eta) \xi_1(\eta).
\label{U-of-tau-m}
\ee
Now we need to estimate the last term (integral) in this equation. To this end let us use again Eq.~(\ref{new_conserved}) but now take the upper integration limit equal to the moment when $\xi$ happens to be at the minimum of $U$, at the nearest point from $\tau_m$ taken above, so that $\tau = \tau_a$. According to Eq.~(\ref{eq:int_source_potential}) the potential terms and the integral of $z' \xi_a$ cancel out and we are left with
\be\label{max_kin_estim}
 \int^{\tau_a}_{\tau_{a0}}   d\eta\,z'(\eta) \,\xi_1 (\eta)  =  \frac{1}{2}\,[\xi'(\tau_a)]^2 -  \frac{1}{2}\,[\xi'(\tau_{a0})]^2 .
\ee
When $\xi$ passes through the minimum of the potential its velocity reaches maximum value for a given oscillation
and $[\xi (\tau_a)']^2  = \Omega^2 \alpha^2 $, where $\alpha$ and $\Omega$ are given by~\rif{eq:xi_amplitude_solution} and~\rif{eq:frequency_2}
respectively. Notice that $[\xi (\tau_a)']^2 $ rises with time and thus for large time  $[\xi'(\tau_{a0})]^2$ may be neglected.
It worth noting that in the limit of harmonic oscillations symmetric with respect to the minimum of $U$ the integral in the
r.h.s. of Eq.~(\ref{max_kin_estim}) does not rise with time but, as we have seen above, the oscillations are
strongly asymmetric with respect to $\xi_a$ and because of that the integral rises with time and Eq.~(\ref{max_kin_estim})
is self-consistent.

The upper integration limits in Eqs.~(\ref{U-of-tau-m}) and (\ref{max_kin_estim}) are slightly different: $\tau_m$ corresponds to the moment
when $\xi$ reaches its minimum value (since $\xi<0$, maximum in absolute value), while $\tau_a$ is the nearest time moment when $\xi$ passes through
the minimum of the potential. So they differ by about a quarter of a period as calculated when $\xi$ is on the left side of the potential minimum.
Since this time interval is quite short the difference between these two integral can be neglected.

So we finally obtain
\be
U(\tau_m) = U_a(\tau_m) + \frac{1}{2} [\xi'(\tau_a)]^2 \approx \frac{1}{2}\left (\alpha^2\Omega^2-\frac{1}{n\,z^{2n}}\right).
\label{U-of-tau-m2}
\ee
Using expression (\ref{eq:potential_+_-}) for $U_- (\xi)$ and relation between $\xi$ and $y$: $y\simeq -\xi/g$, we find
\be
y^2 - 2zy - \frac{1}{g}\left[\alpha^2\Omega^2-\frac{1}{n\,z^{2n}}\right] = 0\,,
\ee
which yields the amplitude
\be\label{eq:beta_spikes}
\beta_{spikes} = \sqrt{\frac{1}{g}\left(\alpha^2\Omega^2-\frac{1}{n\,z^{2n}}\right)+z^2}\,.
\ee
Quite remarkably, this result is exactly equivalent to~\rif{eq:y_amplitude_evolution} in the limit $gz^{2n+2}\gg 1$, so we will assume that from the moment the harmonic approximation fails to be accurate, $\beta$ will follow~\rif{eq:beta_spikes} up to the asymptotic harmonic regime where $gz^{2n+2}\gg 1$. In particular, the moment of transition from harmonic to spike regime is roughly the time at which
\be
\beta_{harm} = \beta_{spikes}\,.
\ee


\section{Gravitational Particle Production}
As is well known, an oscillating curvature gives rise to gravitational particle production. Basically, the energy stored in oscillating gravitational degrees of freedom is released into pairs of elementary particles/antiparticles. As shown in~\cite{Arb_Dolg_Rev} in the case of a minimally-coupled scalar field, the energy released into particles per unit volume and unit time is
\be\label{eq:PP_harmonic}
\dot\rho_{PP}\simeq \frac{\Delta_R^2\,\omega}{1152\pi}\,,
\ee
where $\Delta_R$ is the amplitude of the oscillations of $R$, and $\omega$ is their (physical) frequency. In our case,
\be
\Delta_R \equiv \beta\,T_0\,.
\ee
Moreover, the lifetime of oscillations of $R$ is
\be\label{eq:lifetime}
\tau_R = \frac{48\,\mpl^2}{\omega^3}\,.
\ee
Rigorously speaking, this result is valid when the oscillations of $R$ are perfectly harmonic, or at least when $R$ can be separated in a slowly-varying and an oscillating part, which is supposed to have a constant (or almost constant) frequency. As we have seen in the previous section, solutions of Eq. \rif{eq:xi_evol} show spikes, which are very far from being harmonic oscillations. Thus, we must find a more general result than \rif{eq:PP_harmonic}.

We consider the gravitational particle production of pairs of massless scalar particles $\varphi$, quantized in the usual way:
\be
\begin{gathered}
\hat\varphi(x)= \int \frac{d^3 k}{(2\pi)^32 E_k}\left[\hat a_k\,e^{-ik\cdot x}+\hat a_k^\dagger\,e^{ik\cdot x}\right]\,,\qquad (x\cdot k\equiv \omega t-\mb k\cdot \mb x)\\
\left[\hat a_k,\hat a_{k'}^\dagger\right] = (2\pi)^3\,2E_k\,\delta^{(3)}\left(\mb k-\mb{k'}\right)\,.
\end{gathered}
\ee
At first order in perturbation theory, the amplitude for the creation of two particles of 4-momenta $p_1$ and $p_2$ is equal to~\cite{Arb_Dolg_Rev}:
\begin{align}
A(p_1,p_2)&\simeq \frac{1}{6}\int dt\,d^3x\,R(t)\langle p_1,p_2 | \hat\varphi^2 |0\rangle\notag\\
&= \frac{(2\pi)^3}{3\sqrt 2}\,\,\delta^{(3)}\left(\mb p_1+\mb p_2\right)\int dt\,R(t)\,e^{i(E_1+E_2)t}\,.
\label{eq:amplitude_1}
\end{align}
In terms of the Fourier transform of $R$, defined by
\be
R(t) \equiv \frac{1}{2\pi}\int d\omega\,\tilde{\mc R}(\omega)\,e^{-i\omega t}\,,
\label{R-of-t-Fur}
\ee
we can recast Eq. \rif{eq:amplitude_1} as
\be
A(p_1,p_2)\simeq \frac{(2\pi)^3}{3\sqrt 2}\,\,\delta^{(3)}(\mb p_1+\mb p_2)\,\tilde{\mc R}(E_1+E_2)\,.
\label{A-of-p-2}
\ee
In order to calculate the number of particles produced per unit time and unit volume, we need to integrate $|A(p_1,p_2)|^2$ over all phase space and to divide by the 3-dimensional volume and time duration of the process, $\Delta t$. This yields
\be\label{eq:n_PP}
\dot n_{PP}\simeq \frac{1}{288\pi^2\Delta t}\int d\omega \left|\tilde{\mc R}(\omega)\right|^2\,,
\ee
{and correspondingly, because each particle is produced with energy $E=\omega/2$,}
\be\label{eq:rho_PP}
\dot\rho\simeq \frac{1}{576\pi^2\Delta t}\int d\omega\,\omega\left|\tilde{\mc R}(\omega)\right|^2\,.
\ee
The time duration of the process $\Delta t$ may be considered infinitely large if the characteristic frequency satisfies
the condition $\omega_{ch} \Delta t \gg 1$. Correspondingly, the square of the delta functions in $\tilde{\mc R}$ would be proportional 
to $\delta (0) \sim \Delta t$. For instance, when $R$ is a perfect sine or cosine of frequency $\omega$, one has 
\be\label{eq:deltas}
\tilde{\mc R}(\varepsilon) \sim \delta(\varepsilon-\omega)+\delta(\varepsilon+\omega)\,,
\ee 
and
\be
\left|\delta(\varepsilon\pm \omega)\right|^2 = \frac{\Delta t}{2\pi}\,\delta(\varepsilon\pm \omega)\,.
\ee
Thus $\Delta t$ does not appear in the probability of particle production per unit time.
The physical cut-off of $\Delta t$ in the considered case is given roughly by $\tau_R$ in Eq.~(\ref{eq:lifetime}), therefore for $\omega \gg 1/\tau_R$ the approximation used here is accurate enough. Moreover, since frequencies must be positive, only the first delta function in~\rif{eq:deltas} gives a non-vanishing contribution.

\subsection{Regular Region}
First, let us concentrate on the ``regular'' region (see subsections 3.1 and 3.2). 
Substituting $\Delta_R\equiv \beta\,T_0$ in~\rif{eq:PP_harmonic}, using~\rif{eq:y_amplitude_evolution}
 and \rif{eq:Omega}, and taking $y'_0\simeq 0$ for simplicity, we find
\be
\dot \rho_{PP,\,reg} = \frac{\pi \sqrt{6 n} }{18}\, \frac{t_U \rho_c^{n+1}}{m_{Pl}^4 t_{contr}^2 \rho_{m0}^{n-1}}\,
\frac{ \left( 2n+1+g \right)^{5/2} }{\left(\frac{2n+1}{z^{2n+2}} + g \right)^2} =
C \left(\frac{t_U}{t_{contr}}\right)^2\, \frac{ \rho_c^2}{m_{Pl}^4 t_U}, 
\label{dot-rho-reg}
\ee
where coefficient $C$ has the following  expressions in the two limits:\\
1) $g z^{2n +2} < 1$: 
\be
C =  \frac{\pi \sqrt{ 6 n (2n+1) } }{18} \left( \frac{\rho_c}{\rho_{m0} }\right)^{n-1}\, z^{4n+4} ;
\label{C-small-g}
\ee
2) $g z^{2n +2} > 1$:
\be
C = \frac{ \pi  \left[ 6 n (2n+1+g)  \right]^{5/2} }{18} \left( \frac{\rho_c}{\rho_{m0}} \right)^{5n+3}\,(m t_U)^4.
\label{C-large-gz}
\ee
The last factor in Eq.~(\ref{dot-rho-reg}) is extremely small. Since $\rho_c^2 \sim m^4_{Pl}/ t_U^4$, this factor
is about $1/t_U^5$. So unless $C$ is very large, particle production in the regular region would be negligible.
The most favorable possibility would be small $g$ and large $z$, but keep in mind that $g \sim \rho_{m0}^{2n+2}$.
We present an estimate of the flux in the conventional units as:
\be \label{eq:part_prod_rate}
\frac{\dot\rho_{PP,\,reg}}{\text{GeV\,s}^{-1}\text{m}^{-3}} \simeq 3.6\times 10^{-141}\, \frac{C_1(n,g,z)\,z^{4n+4}}{\rho_{29}^{n-1} t_{10}^2 }
\simeq  2.5\times 10^{47}\, \frac{C_2 (n,g,z)\,m_5^4}{\rho_{29}^{5n+3} t_{10}^2}.
\ee
The coefficients $C_1$ and $C_2$ are convenient to use when $gz^{2n+2}\ll 1$ and $gz^{2n+2}\gg 1$. 
They are respectively:
\begin{subequations}\label{C1-C2}
\begin{align} 
C_1 &= \sqrt{n(2n+1 +g)} \left(\frac{2n+1+g}{2n+1 + g z^{2n+2}}\right)^2 \approx \sqrt{n(2n+1)},  \\
C_2 &= n^{5/2} \sqrt{2n+1+g}\,g^2 \left(\frac{2n+1+g}{2n+1 + g z^{2n+2}}\right)^2 z^{4n+4}\approx {[n(2n+1)]^{5/2}}.
\end{align}
\end{subequations}

\subsection{Spike Region}
In the spike region the particle production rate would be strongly enhanced due to much larger amplitude of the
oscillations of $R$. We parametrize the solutions in this region
as a sum of gaussians with slowly varying amplitude $\Delta_R(t)$, 
superimposed on the smooth power-like background, $-T(t)$:
\be\label{eq:gaussians}
R(t) = -T(t)+\Delta_R(t)\sum_{j=1}^{N}\exp\left[-\frac{(t-jt_1)^2}{2\sigma^2}\right]\,.
\ee
Here $t_1$ is the time-shift between the spikes and $\sigma$ is the width of the spikes. 
The values of these parameters are determined from the solution obtained above.
The slow variation of the functions $T(t)$ and $\Delta_R(t)$ means that  $\dot T\ll T/t_1$ and $\dot \Delta_R\ll \Delta_R/t_1$.\\
In principle, $N$ could be infinitely large, which for $\Delta_R =$ const. corresponds to an infinitely long duration of the process. As we have mentioned before, this does not have an essential impact on the probability of particle production per unit time.

In accordance with the solutions of the equations determining the evolution of curvature, we consider the case 
$\sigma\ll t_1 $, that is the spacing between the spikes is much larger than their width. 
At high frequencies the Fourier transform of \rif{eq:gaussians} is dominated by the contribution of the quickly-varying gaussians, i.e.
\be
\tilde{\mc R}(\omega)\simeq \sqrt{2\pi}\,\Delta_R|\sigma|e^{-\omega^2\sigma^2/2+i\omega t_1}\,\,\frac{e^{iN\omega t_1}-1}{e^{i\omega t_1}-1}\,.
\ee
When squared, this gives
\be\label{eq:tilde_R_2}
\left|\tilde{\mc R}(\omega)\right|^2 \simeq 2\pi \Delta_R^2\,\sigma^2e^{-\omega^2\sigma^2}\,\frac{\sin^2 N\omega t_1/2}{\sin^2 \omega t_1/2}
\ee
The dominant part of this expression comes from $\omega=\omega_j = 2j \pi/t_1$, where it is equal to $N^2$. Around these points we have
\be
\frac{\sin^2 N\omega t_1/2}{\sin^2\omega t_1/2}\underset{\omega\simeq 2j\pi/t_1}{\simeq} \left(\frac{\sin \left[N(\omega t_1/2-j\pi)\right]}{\omega t_1/2-j\pi}\right)^2\,.
\ee
We take the limit $N\to\infty$ and use a representation of Dirac's delta function to write
\be\label{eq:sin2}
\frac{\sin^2 N\omega t_1/2}{\sin^2\omega t_1/2} \simeq \sum_j\left|\pi\,\delta\left(\frac{\omega t_1}{2}-j\pi\right)\right|^2 = 
\sum_j\left|\frac{2\pi}{t_1}\,\delta\left(\omega-\omega_j\right)\right|^2\,,
\ee
which yields
\be
\left|\tilde{\mc R}(\omega)\right|^2\simeq \frac{4\pi^2 \Delta_R^2\,\sigma^2e^{-\omega^2\sigma^2}\Delta t}{t_1^2}\sum_j\delta\left(\omega-\frac{2\pi j}{t_1}\right)\,.
\ee
Each particle produced by the oscillation component of frequency $\omega$ has energy $\omega/2$, 
so the gravitational particle production rate is
\begin{align}\label{eq:PP_general}
\notag \dot\rho_{PP} &= \frac{1}{288\pi^2\Delta t}\int\,d\omega\,\frac{\omega}{2}\left|\tilde{\mc R}(\omega)\right|^2\\
 \notag &= \frac{\pi \Delta_R^2\,\sigma^2}{72\,t_1^3}\sum_{j} j\exp\left[-\left(\frac{2\pi j\,\sigma}{t_1}\right)^2\right]\\
&\simeq \frac{\Delta_R^2}{576\pi\,t_1}\,.
\end{align}
In the last step, we have again assumed that $\sigma/t_1\ll 1$, so that the summation over $j$ can be replaced by an integral. Remarkably, the dependence on $\sigma$ has disappeared from the result.
Lastly, we use $\Delta_R=\beta\,T_0$ and~\rif{eq:beta_spikes}, obtaining
\be
\dot \rho_{PP,\,sp} = \frac{ (6 n)^{3/2} (2n+1+g)^{5/2} }{ 18 (2n+1 + gz^{2n+2}) }\,\frac{ m^2 t_U^3 z^{2n+2} }{t_{contr}^2 m_{Pl}^4}
\frac{ \rho_c^{3n+3} }{\rho_{m0}^{3n+1}}.
\label{dot-rho-sp}
\ee
or in conventional units:
\be\label{eq:part_prod_spikes}
\frac{\dot\rho_{PP,\,sp}}{\text{GeV\,s}^{-1}\text{m}^{-3}} \simeq 3.0 \times 10^{-47}\,\frac{C_3(n,g,z)\,m_5^2\,z^{2n+2}}{t_{10}^2\,\rho_{29}^{3n+1}}
\ee
where
\be
C_3 = \frac{(2n+1+g)^{5/2}n^{3/2}}{(2n+1+gz^{2n+2})} \mb{\approx [n(2n+1)]^{3/2}}\,.
\label{C3}
\ee
All elementary particles couple to gravity, so in order to get an order-of-magnitude estimate of the overall particle production 
one should multiply \rif{eq:part_prod_rate} and \rif{eq:part_prod_spikes} by the number of elementary particle species,  $N_s$,
with masses bound from above by $m \lesssim  2\pi/\sigma $.

\subsection{Backreaction on Curvature and Mode-dependent Damping}
So far we have not taken into account that the oscillation amplitude should be damped due to the back reaction of particle production. Neglecting such damping 
would be an accurate approximation up to $t/\tau_R \simeq 1$; for larger times, however, the damping should be
taken into consideration. In the regular region oscillations are practically harmonic, so only a single
frequency mode is involved and one simply needs to add the exponential damping 
factor $\exp[-2\Gamma(\omega)t]$ to \rif{eq:PP_harmonic} and \rif{eq:part_prod_rate}. In the spike region the problem is more complicated because, according to Eq.~(\ref{eq:lifetime}), the damping depends upon the frequency, so different modes are  damped differently and this can noticeably distort the form of the initial $R(t)$.
A simple approximate way to take into account this damping is to introduce the factor $\exp [-t/\tau_R (\omega)] $ into the integrand of Eq.~(\ref{R-of-t-Fur}). After a sufficiently long time, only the modes with the lowest frequency survive and one may naively expect that the lowest frequency modes give the dominant contribution to particle production. However, one should keep in mind that the time duration is finite: in fact, it is equal to the time of stabilization of the collapsing system and is surely shorter than the cosmological time $t_c \approx 4\cdot 10^{17} $~s. Thus the the energy is predominantly  emitted with the frequencies determined by the condition  $\Delta t /\tau_R(\omega)\simeq 1 $, see discussion 
below Eq.~(\ref{rho-fin}).

So to take into account the damping of $R$-oscillations we introduce into the amplitude (\ref{eq:amplitude_1}) 
the damping factor $\exp [-\Gamma (\omega) t ]$, where $\Gamma (\omega) = 1/\tau_R(\omega) = \omega^3/48 m_{Pl}^2$ and integrate
over time up to a finite upper limit:
\bea
A(p_1,p_2) &\simeq & \frac{(2\pi)^3}{3\sqrt 2}\,\,\delta^{(3)}\left(\mb p_1+\mb p_2\right)
\int \frac{d\omega}{2\pi} \, \tilde{ R}(\omega) \int_0^{t} dt' \, e^{i(E_1+E_2 - \omega)t' -\Gamma(\omega) t'} \nonumber \\
&=& \frac{(2\pi)^3}{3\sqrt 2}\,\,\delta^{(3)}\left(\mb p_1+\mb p_2\right) 
\int \frac{d\omega}{2\pi} \, \tilde{ R}(\omega)\, 
\frac{\exp[ i(2E -\omega)t - \Gamma(\omega) t]-1}{i ( 2E - \omega) -\Gamma(\omega)}\,,
\label{A-damped}
\eea
where $E= E_1 = E_2$ {and $t$ has here the same meaning as $\Delta t$ in \rif{eq:n_PP}}. 

As we have seen above, $\tilde R (\omega) $ can be written as:
\be
\tilde R (\omega) \simeq \sqrt{2}\,\pi^{3/2} \Delta_R\,\sigma e^{-\omega^2 \sigma^2/2}\, e^{i\Phi} 
\sum_j \delta\left(\frac{\omega t_1}{2} - \pi j \right),
\label{tilde-R}
\ee
where $\exp (i\Phi)$ is a phase factor of modulus unity. Substituting this expression in Eq.~(\ref{A-damped}) and integrating over $\omega$, we find up to a phase factor:
\be
A(p_1,p_2) \simeq  \frac{(2\pi)^{7/2} \Delta_R\,\sigma }{3\sqrt 2\, t_1}\,\,\delta^{(3)}\left(\mb p_1+\mb p_2\right)
\sum_j e^{-\sigma^2\omega_j^2/2}\,\frac{1-\exp\left[(i \epsilon_j -\Gamma_j)t\right]}{\epsilon_j +i\,\Gamma_j}  ,
\label{A-fin}
\ee
where $\omega_j = 2\pi j /t_1$, $\epsilon_j = 2E - \omega_j$, and $ \Gamma_j = \Gamma (\omega_j)$.

The energy density of the produced particles is:
\begin{subequations}\label{rho-damped}
\begin{align}
\rho &= \frac{\Delta_R^2\,\sigma^2 }{72\pi t_1^2}\,\int_0^\infty dE\,E
\sum_j \,e^{-\sigma^2 \omega_j^2}\,\frac{1 + e^{-2\Gamma_jt} -2\,e^{-\Gamma_jt} \cos (\epsilon_jt)}{\epsilon_j^2 +\Gamma_j^2} \\
&=\frac{\Delta_R^2\,\sigma^2}{288\pi t_1^2}\sum_j e^{-\sigma^2\omega_j^2}
\int_{-\omega_j t}^\infty\,d\eta_j(\eta_j+\omega_j t)\,\frac{1 + e^{-2\Gamma_jt} -2\,e^{-\Gamma_jt} \cos \eta_j}{\eta_j^2 +(\Gamma_jt)^2}\,. 
\end{align}
\end{subequations}
We have introduced here  a new integration variable
$\eta_j\equiv \epsilon_j t$. It was also assumed that the diagonal terms with $j=k$ dominate in the double sum over $j$ and $k$ in the expression for $|A (p_1,p_2)|^2$; this is a good approximation for small $\Gamma$, in particular, $\Gamma \ll \omega$. 
Assuming $\omega_jt\gg 1$, the integral is easily taken at the poles of the denominator and we finally obtain:
\be
\rho = \frac{\pi \Delta_R\, \sigma^2}{144\,t_1^3} \,\sum_j j\,e^{-\sigma^2\omega_j^2}\,\frac{1 - e^{-2\Gamma_j t}}{\Gamma_j}.
\label{rho-fin}
\ee
For $\Gamma_j t \ll 1$ this result coincides  with (\ref{eq:PP_general}) after dividing by the total elapsed time $t$. The summation over $j$ can be separated into two regions of large and small $\Gamma_j t$. The boundary value of $\omega_j$ is given by the condition $2\Gamma_b t=\omega_b^3 t /24 m_{Pl}^2 =1$. Correspondingly
\be
\omega_b \simeq  {180} \,{\rm MeV} \left( t_c / t\right)^{1/3}\,,
\label{omega-b}
\ee
where $t_c = {4}\cdot 10^{17}$ s is the cosmological time. The boundary value of $j$ is 
\be
j_b  = \frac{ \omega_b t_1 }{2\pi} = \frac{\omega_b}{ \Omega m \sqrt{g}}, 
\label{j-b}
\ee
where we took $t_1 = 2\pi/\omega$ with  $\omega$ defined in eqs.~\rif{eq:frequency_2} and \rif{eq:Omega}.
In particular, for small $g$ we have $\Omega \simeq z^{n+1}/\sqrt{2n+1}$ and using Eq.~\rif{g-kappa} we find
\be
j_b \simeq {1.6} \cdot 10^{41}\, \frac{\sqrt{(2n+1) {n}}}{(z \rho_{29})^{n+1}}\,\left(\frac{t_c}{t}\right)^{1/3}.
\label{j-b2}
\ee

Accordingly, the time $t_1$ can be estimated as $t_1 \approx 4\cdot 10^{18}\,{\rm  s}\, \sqrt{(2n+1)n}/(z \rho_{29})^{n+1}$.\\
Separating the summation over $j$ into two intervals of small and large $\Gamma_j t$, we obtain:
\be\label{eq:rho-sum-separ}
{\rho \approx \frac{\pi \Delta_R^2\,\sigma^2}{144\,t_1^3} \,\left[ 2t \sum_{j=1}^{j_b}  j\,e^{-(\sigma \omega_j)^2} + \sum_{j=j_b}^{\infty} 
 \frac{j\,e^{-(\sigma \omega_j)^2}}{\Gamma_j}\right]\,.}
\ee
If $\sigma \sim 1/m$, the exponential suppression factor $\exp (-\sigma \omega_j)^2$ is weak near $j = j_b$ and the sums over $j$ can be easily evaluated:
\be
{\rho \approx \frac{\pi \Delta_R^2 \sigma^2}{144\,t_1^3} \,\left( j_b^2 t + \frac{6 m_{Pl}^2 t_1^3}{\pi^3 j_b}\right) }\,.
\label{rho-fin-sum}
\ee
Alternatively, using the fact that $\sigma/t_1\ll 1$, we can replace the summation with an integral, obtaining the similar, more general result:
\begin{align}
\rho &\simeq \frac{\pi \Delta_R^2\,\sigma^2}{144\,t_1^3}\left[2t\int_0^{j_b}dj\,j\,e^{-\sigma^2\omega_j^2}+\int_{j_b}^\infty dj\,\frac{j\,e^{-\sigma^2\omega_j^2}}{\Gamma_j}\right] \notag\\
&\simeq \frac{\Delta_R^2\,t}{576\pi\,t_1}\left[1-e^{-\sigma^2\omega_b^2}\right]+\frac{\Delta_R^2\,\mpl^2\,\sigma^2}{12\pi\,\omega_b\,t_1}\left[e^{-\sigma^2\omega_b^2}-\sqrt\pi\,\sigma \omega_b\,\text{erfc}(\sigma\omega_b)\right]\,,
\label{eq:rho-fin-integral}
\end{align}
where $\text{erfc}(x)$ is the complementary error function:
\[
 \text{erfc}(x) = \frac{2}{\sqrt\pi}\int_x^{\infty} dt\,e^{-t^2}\,.
\]
Note that for $\sigma\omega_b\ll 1$, as was assumed before, this gives
\be
\rho\simeq \frac{\Delta_R^2\,\omega_b^2\,\sigma^2}{576\pi\,t_1}\left(t+\frac{48\mpl^2}{\omega_b^3}\right)= \frac{\Delta_R^2\,\omega_b^2\,\sigma^2\,t}{192\pi\,t_1}\,,
\label{eq:rho-fin-sum-equiv}
\ee
which is exactly equivalent to \rif{rho-fin-sum}. For $\sigma\omega_b\gg 1$, instead, we recover \rif{eq:PP_general}. 
This makes sense, because $\sigma\omega_b\gg 1$ corresponds to $j_b\to\infty$ and hence to the limit in which the time elapsed is not long enough for particle production to have had a noticeable back-reaction on curvature. Nonetheless, during this time particles may have been effectively produced.

These estimates are valid even in the spike region when $g z^{2n+2} \ll 1$ but $gy^{2n+2}$ may reach values much larger than unity.

\subsection{Damping of Oscillations \label{ss-damping}}
As we have seen, having a wide frequency spectrum of 
the Fourier transform of $R$ makes it impossible to simply use an exponential damping $R\to R\,e^{-\Gamma t}$ to include the effects of particle production. Moreover, the increasing energy density acts as a source term and increases the amplitude of the oscillations of $R$, which makes things even more complicated. However, the picture for the field $\xi$ is relatively simple, because its oscillations are almost harmonic. We still have a source component, given by the increasing $z$, but we can once again use the energy conservation equation to determine the time at which oscillations basically stop due to the damping. The effect of particle production on the evolution equation for $\xi$ (see Eq.~\ref{eq:xi_evol_approx}) is to transform it into
\be
\xi_1'' + 2\gamma(\Omega)\,\xi_1' + \Omega^2\xi_1 = -\xi_a''\,,
\ee
where
\be
\gamma(\Omega)\equiv \frac{\Gamma(m\,\Omega \sqrt g)}{m\sqrt g} = \frac{\Omega^3m^2g}{48\,\mpl^2}\,,
\ee
which in fact for $\gamma/\Omega\ll 1$ and $\xi_a''/\xi_1''\ll 1$ generates the wanted behaviour
\be
\xi_1 \sim e^{-\gamma\tau}\,\sin\Omega\tau\,.
\ee
Correspondingly, the energy conservation~\rif{eq:conserved} becomes
\be
\frac{1}{2}\,\xi'^2 + U(\xi) + 2\int_{\tau_0}^\tau d\eta\,\gamma\,\xi_1'^2 - \int^\tau_{\tau_0} d\eta \,z'\,\xi 
= \frac{1}{2}\,\xi_0'^2+ U(\xi_0) = \text{ const. }  
\label{balance-gamma}
\ee
Let us consider values of $\tau$ in which $\xi(\tau) = \xi_a(\tau)$. In this case the potential disappears from this 
equation [see Eq.~\rif{eq:int_source_potential}], which can be now symbolically written as an equality between 
the variation of the kinetic energy and two integral terms:
\begin{subequations}\label{eq:balance_PP}
\be
\Delta K = I_{source} - I_\gamma\,,
\ee
where
\be
\Delta K = K-K_0 \equiv \frac{1}{2}(\xi'^2 - \xi_0'^2)\,,\qquad I_{source} \equiv \int^\tau_{\tau_0}d\eta\,z'\,\xi_1\,,
\qquad I_\gamma \equiv 2\int^\tau_{\tau_0}d\eta\,\gamma\,\xi_1'^2\,.
\ee
\end{subequations}
By definition $\xi = \xi_a$ is the position of $\xi$ at the minimum of the potential. So when
$\xi=\xi_a$, the kinetic energy, $K$, reaches one of the local maxima (in time).
This picture is particularly clear if we compare the system with a classical oscillator: when the field passes through the equilibrium point, the potential has minimal value and the velocity is maximal.

We estimate the effect of damping perturbatively applying the energy balance (\ref{balance-gamma})
with unperturbed functions $\xi (\tau)$ for which the effects of damping are neglected. In the absence of damping, the condition (\ref{eq:balance_PP}) turns into $\Delta K = I_{source}$, which is essentially Eq. (\ref{max_kin_estim}). Evidently the impact of damping on the oscillations of $\xi$ starts to be important when $I_\gamma$ becomes of the order of $\Delta K$. Though the damping coefficient is small, \b{i.e.} $\Gamma \ll \omega$, the integral $I_\gamma$ rises with time faster than $\Delta K$ and ultimately it will overtake it. So we need to check when the condition 
\be 
\alpha^2 (\tau)\Omega^2 (\tau)  \sim 2\int^\tau d\eta\, \gamma (\eta) \alpha^2 (\eta) \Omega^2 (\eta)
\label{K-equal-gamma}
\ee
starts to be fulfilled with $\alpha$ and $\Omega$ given by~\rif{eq:xi_amplitude_solution} and~\rif{eq:frequency_2}.

Keeping in mind that $\gamma \tau = \Gamma t$ and using Eq.~(\ref{eq:lifetime}) for $\Gamma = 1/\tau_R$
with $\omega  = \Omega m \sqrt{g}$, we find that the equality~\rif{K-equal-gamma} is satisfied when the energy density is equal to
\be
\begin{aligned}
z_\gamma^{3n+4} &= \frac{24(2n+1)^{3/2}(4n+5)\kappa\,\mpl^2}{g\,m^2}\\
&= \frac{24(2n+1)^{3/2}(4n+5)}{g^{3/2}}\left(\frac{\mpl}{m}\right)^3\frac{1}{\mpl\,t_{contr}}\\
&\simeq 6\times 10^{123}\,\frac{[n(2n+1)]^{3/2}(4n+5)}{t_{10}\,\rho_{29}^{3n+3}}
\end{aligned}
\label{t-gamma}
\ee
The corresponding boundary value of $t$ is $t_\gamma = z_\gamma\,t_{contr}$. 

For times smaller than $t_\gamma$ the effects of damping are negligible and particle production 
can be very effective giving rise to substantial production of cosmic rays, as we shall see in the next section.
Particle production is especially pronounced in the spike region when the amplitude of $R$ is very large.

As we have seen, the spikes' width is $\sigma\sim m^{-1}$ while their spacing, which determines the effective frequency, is $t_1\sim \omega^{-1}$. High frequency oscillations of $R$ should be damped very rapidly, since $\Gamma \sim \omega^3$, but  due to the non-harmonicity of the potential and the non-linearity of the relation between $\xi$ and $y$ or $R$, the low frequency oscillations are efficiently transformed into high frequency spikes of small amplitude in $\xi$ but of very large amplitude in $R$. It is worth noting in this connection that the bulk of the energy density associated with the oscillations of $R$ is concentrated at low frequencies, $\rho_{osc} \sim \Delta_R^2 m_{Pl}^2 /\omega^2 $, so the energy reservoir at low frequencies is deep enough to feed up the spikes.

The physical frequency, $\omega= \Omega m \sqrt{g}$, depends upon the product $Q \equiv g z^{2n+2}$. If $Q \ll 1$ the frequency may be rather low, of the order of hundred MeV, while for $Q \gg 1$ the frequency reaches 
the maximum value $\omega = m$. In the first case the lifetime of harmonic oscillations could be larger than
the universe age, while in the second case  it would be shorter than a second.

\section{Estimate of Cosmic Ray Emission}
\subsection{Regular Region}

Let us consider a cloud (e.g. a protogalaxy) with total mass $M$ and density $\rho$. Particles
would be uniformly produced over its whole volume, which is equal to:
\be\label{eq:volume}
V=\frac{M}{\rho} = 2\times 10^{73} \,{\rm cm}^3 \,\frac{M_{11}}{z\,\rho_{29}}\,,
\ee
where the mass of the cloud, $M$, is expressed in terms of the solar mass $M_\sun$:
\be
M_{11}\equiv \frac{M}{10^{11}M_\sun} = \frac{M}{2\times 10^{44}\text{ g}}\,.
\ee
In the regular case the oscillations of $R$ are almost harmonic, so we can rely on the adiabatic approximation and use Eq.~(\ref{eq:PP_harmonic}) 
for the particle production rate or Eq.~(\ref{eq:part_prod_rate}) 
corrected by the damping factor ${\exp[-2\Gamma (\omega) t]}$. To be more precise, in the exponent we should take the integral 
of $\Gamma$ over time to take into account the (slow) variation of $\omega$.

The total luminosity relative to gravitational particle production is obtained multiplying the rate of energy production per unit time and 
volume \rif{eq:part_prod_rate} by the total volume \rif{eq:volume}, that is $L= V(t)\dot\rho_{PP}(t)$, or
\be\label{eq:luminosity_regular}
\frac{L_{reg}}{\text{GeV s}^{-1}}\simeq 7.3 \times 10^{-74}\,N_s\,\frac{\sqrt{(2n+1+g)n}\,M_{11}\,z^{4n+3}}
{\left(1+\cfrac{gz^{2n+2}}{2n+1}\right)^2\rho_{29}^n\,t_{10}^2}\, \exp \left[-2\int^t_{t_0}  dt'\,\Gamma (\omega) \right]\,,
\ee
where $\omega$ depends upon time due to the variation of $z(t)$, see eqs.~\rif{eq:frequency_2}, \rif{eq:Omega}. The initial time $t_0$ should be taken at the onset of structure formation, when the energy density locally started to rise.

For $gz^{2n+2}\ll 1$, the luminosity is negligible with respect to the luminosity
in the spike region (see below), so we will not consider this case further. When $gz^{2n+2}$ becomes larger than unity and $\omega \sim m$, the lifetime of oscillations turns out to be at most a few seconds and an explosively fast particle production takes place. The integrated luminosity can be approximately obtained from Eq.~(\ref{eq:luminosity_regular}) dividing it by $\Gamma (m)$ and taking the exponential factor equal to 1. 
The time duration of the production process, of the order of a few seconds, is close to that of some Gamma Ray Bursts but the characteristic particle energies are much higher, instead of MeV it is of the order of the scalaron mass $m\gtrsim 10^5$ GeV.

If $g>1$, oscillations are very mildly excited and particle production is negligible. This can be seen from \rif{eq:y_amplitude_evolution} with $g>1$, keeping in mind that $\kappa\sim g^{-1/2}$. Alternatively one may use Eq.  \rif{eq:luminosity_regular}
with  $g>1$ taking into account that a large value of $g$ corresponds to large values of $\rho_{29}$ and/or of $n$.
 
\subsection{Spike Region}
There remains to consider the spike region, where we need to use~\rif{eq:part_prod_spikes} instead of \rif{eq:part_prod_rate}, or \rif{rho-fin} and \rif{rho-fin-sum}. Equation \rif{eq:part_prod_spikes} yields
\be
\label{eq:luminosity_spike}
\frac{L_{sp}}{\text{GeV s}^{-1}} \simeq 6.0\times 10^{20}\,\frac{C_3\,N_s\,M_{11}\,m_5^2\,z^{2n+1}}{t_{10}^2\,\rho_{29}^{3n+2}}
\ee
where $C_3$ is given by Eq.~(\ref{C3}). 
This result is valid when $\Gamma_j t < 1$ for all essential values of $j$, see Eq.~\rif{rho-fin}, that is when the damping due to particle production is negligible. See also the discussion in Sec. \ref{ss-damping}.

In the opposite case, we cannot use the approximate account of damping made above (see Eq.~\ref{eq:luminosity_regular}) 
because oscillations are strongly anharmonic. If modes with both $\Gamma_j t$ greater and smaller than unity are essential, we 
have to use eqs.~\rif{rho-fin} or {\rif{eq:rho-fin-integral}}.
We start by rewriting Eq.~\rif{eq:rho-fin-sum-equiv} using $\omega_b^3=24\mpl^2/t $, $\Delta_R=\beta_{spikes}\,T_0$ and $t_1=2\pi/\omega$, where $\beta_{spikes}$ and $\omega$ are given, respectively, by \rif{eq:beta_spikes} and \rif{eq:frequency_2}-\rif{eq:Omega}. This yields, assuming $gz^{2n+2}\ll 1$ and $t\simeq z\,t_{contr}$,
\be
\frac{\rho}{\text{GeV m}^{-3}} \simeq 1.1\times 10^{-40}\,\frac{C_3\,(\sigma m)^2\,z^{2n+7/3}}{t_{10}^{5/3}\,\rho_{29}^{3n+1}}
\ee
and
\be\label{eq:spikes_rhodot_damped}
\frac{\dot\rho}{\text{GeV\,s}^{-1}\text{m}^{-3}} \simeq 6.9\times 10^{-58}\,\frac{\tilde C_3\,(\sigma m)^2\,z^{2n+4/3}}{t_{10}^{8/3}\,\rho_{29}^{3n+1}}\,,
\ee
where
\be
\tilde C_3 = \frac{(2n+1)(n+1)}{(2n+1+gz^{2n+2})}\,C_3\,.
\ee
From~\rif{rho-fin} it is clear that for each mode labelled by $j$ we have
\be
\dot\rho_j \simeq \frac{\pi\,\Delta_R^2\,\sigma^2j\,e^{-\sigma^2\omega_j^2}}{72\,t_1^2}\,e^{-2\Gamma_j t}\,
\ee
which has the predicted behaviour $\dot\rho\sim e^{-2\Gamma t}$. However, this behaviour is not obvious in the complete solution~\rif{eq:spikes_rhodot_damped}. This means that in this case the overall effect is more complicated than a simple exponential damping. The anharmonicity of the oscillations and the dependence of both $\dot\rho$ and $\Gamma$ on the frequency give non-trivial results which were impossible to predict without performing explicit calculations. 

When the damping due to particle production is relevant, the total luminosity becomes
\be
\label{eq:luminosity_spikes_damped}
\frac{L_{sp}}{\text{GeV s}^{-1}} \simeq 1.4\times 10^{10}\,\frac{\tilde C_3\,N_s\,M_{11}(\sigma m)^2z^{2n+1/3}}{t_{10}^{8/3}\,\rho_{29}^{3n+2}}
\ee
This value, though smaller than~\rif{eq:luminosity_spike}, might not be completely negligible, especially for short contraction times and relatively small initial densities. This means that even with the damping of oscillations taken into account, the produced cosmic rays could in principle be detectable.

\section{Discussion and Conclusions}

We have shown that in contracting astrophysical systems with rising energy density, powerful oscillations of curvature scalar $R$ are induced. Initially harmonic, these oscillations evolve to strongly anharmonic ones with high frequency and large amplitude, which could be much larger than the value of curvature in standard General Relativity.

Such oscillations result in efficient particle production in a wide energy range, from a hundred MeV up to the scalaron mass, 
$m$, which could be as large as $10^{10} $ GeV (and maybe even larger). Such high frequency oscillations could be a 
source~\cite{eva-add-lr} of ultra high energy cosmic 
rays (UHECR) with $E \sim  10^{19}-10^{20}$ eV, see e.g. the review~\cite{PDG_2012}, which might avoid the GZK cutoff~\cite{gzk}. 
Possibly the considered mechanism would give too large a fraction of high energy photons in UHECR, see ref.~\cite{UHECR} if no special care is taken, because gravity couples to all elementary particles with the same intensity. However, direct photon production may be suppressed due to the conformal invariance of electrodynamics.
To avoid a too strong indirect photon production one may need to introduce "photo-fobic" heavy particles predominantly created by the oscillating curvature.

It is tempting to explain gamma bursts by these curvature oscillations, but the emitted particle energy seems to be
much above the MeV range. To this end some modification of the model or a mechanism of energy depletion would be necessary, and could be an interesting subject of future research.

The oscillations considered here may also have an essential impact on the gravitational (Jeans) instability in $F(R)$ gravity studied for 
instance in ref.~\cite{capozziello}, where this effect was not taken into consideration.

The efficiency of particle production strongly depends upon the system under scrutiny, the values of the parameters
of the theory, and upon the explicit form of the function $F(R)$. These problems deserve further study, but the framework presented in this paper can be applied to many possible cases.

\subsection*{Acknowledgements} 
We thank D. Gorbunov and G. Rubtsov for the discussion on the chemical composition of ultrahigh energy cosmic rays. EA and AD acknowledge the support of the grant of the Russian Federation government 11.G34.31.0047.




\begin{thebibliography}{99}

\bibitem{Nobel_2011} A.G. Riess et al.,
\textit{Astron. J.}, \b{116} (1998) 1009-1038;\\
S. Perlmutter et al., \textit{Nature} \b{391} (1998) 51;\\ 
B.P. Schmidt et al., \textit{Astrophys. J.} \b{507} (1998) 46;\\
S. Perlmutter et al.,
\textit{Astrophys. J.} \b{517} (1999) 565-586;\\
A.G. Riess et al.,
\textit{Astrophys. J.} {\bf 607} (2004) 665-687.


\bibitem{one-over-R}
S. Capozziello, S. Carloni, A. Troisi,
\textit{Recent Res. Dev. Astron. Astrophys.} {\bf 1} (2003) 625;\\
S.M. Carroll, V. Duvvuri, M. Trodden, M.S. Turner,
\textit{Phys. Rev. D} {\bf 70} (2004) 043528.

 \bibitem{DolgKaw}
A.D. Dolgov, M. Kawasaki, \textit{Phys. Lett. B} {\bf 573} (2003) 1.

 \bibitem{HuSaw}
 W. Hu, I. Sawicki, \textit{Phys. Rev. D} {\bf 76}, 064004 (2007).
 
 \bibitem{ApplBatt}
 A. Appleby, R. Battye, \textit{Phys. Lett. B} {\bf 654} (2007) 7.

\bibitem{Starobinsky_2007}
A.A. Starobinsky, \textit{JETP Lett.} {\bf 86} (2007) 157.

\bibitem{rev-f-of-R}
S.A. Appleby, R.A. Battye, A.A. Starobinsky, \i{JCAP} \b{1006} (2010) 005.

\bibitem{noj-odin}
S. Nojiri, S. Odintsov, \i{Phys. Rept.} \b{505} (2011) 59;\\
K. Bamba, S. Capozziello, S. Nojiri. S.D. Odintsov [arXiv: 1205.3421].

\bibitem{frolov}
A.V.  Frolov,  \textit{Phys. Rev. Lett.} {\bf 101} (2008) 061103;\\
I. Thongkool, M. Sami, R. Gannouji, S. Jhingan, \textit{Phys. Rev. D} {\bf 80} (2009) 043523;\\
I. Thongkool, M. Sami, S. Rai Choudhury, \textit{Phys. Rev. D} {\bf 80} (2009) 127501.
 
 \bibitem{Arb_Dolgov}
E.V. Arbuzova, A.D. Dolgov, \textit{Phys. Lett.} {\bf B} 700, 289 (2011).

\bibitem{bamba-noj-odin}
K. Bamba, S. Nojiri, S.D. Odintsov, \i{Phys. Lett.} \b{B} 698, 451 (2011).

 \bibitem{Gur-Star} 
 V.Ts. Gurovich, A.A. Starobinsky, 
      \textit{Sov. Phys. JETP} \b{50} (1979) 844; [Zh. Eksp. Teor. Fiz. 77 (1979) 1683];\\
      A.A. Starobinsky, \textit{JETP Lett.} \b{30} (1979) 682; 
      [Pisma Zh. Eksp. Teor. Fiz. \b{30} (1979) 719];\\
             A.A. Starobinsky, ``Nonsingular model of the Universe with the 
   quantum-gravitational de Sitter stage and its observational consequences'',
      in \textit{Proc. of the Second Seminar "Quantum Theory of Gravity"} (Moscow,
      13-15 Oct. 1981), INR Press, Moscow , 1982, pp. 58-72; reprinted in:
      Quantum Gravity, eds. M. A. Markov and P. C. West. Plenum Publ. Co.,
      N.Y., 1984, pp. 103-128.

 \bibitem{Starobinsky_1980}
A.A. Starobinsky, \textit{Phys. Lett.} \b{B91}, 99 (1980).

 \bibitem{Zeld-Star}
Ya. B. Zeldovich, A.A. Starobinsky, \textit{JETP Lett.} \b{26} (1977) 252. 

\bibitem{Vilenkin_1985} A. Vilenkin, \textit{Phys. Rev. D} \b{32} (1985) 2511.

\bibitem{Arb_Dolg_Rev}
E.V. Arbuzova, A.D. Dolgov, L. Reverberi, \i{JCAP} \b{02} (2012) 049.

\bibitem{japanese}
H. Motohashi, A. Nishizawa, \i{Phys. Rev. D} {\bf 86} (2012) 083514.


\bibitem{Reverberi2013}
L. Reverberi, \i{Phys. Rev. D} {\bf 87} (2013) 084005.


\bibitem{eva-add-lr}
E.V. Arbuzova, A.D. Dolgov, L. Reverberi, \i{Eur. Phys. J. C} {\bf 72} (2012) 2247.

\bibitem{PDG_2012}
J. Beringer et al. (Particle Data Group), \i{Phys. Rev. D} \b{86} (2012) 01001.

\bibitem{gzk}
K. Greisen, \i{Phys. Rev. Lett.} {\bf 16} (1966) 748;\\
G.T. Zatsepin, V.A. Kuzmin, \i{Sov. Phys. JETP Lett.} {\bf 4} (1966) 064004.


\bibitem{UHECR}
R. Aloisio, V. Berezinsky, M. Kachelriess, \i{Nucl. Phys. Proc. Suppl.} \b{136} (2004) 319;\\
O.E. Kalashev, G.I. Rubtsov, S.V. Troitsky, \i{Phys. Rev. D} \b{80} (2009) 103006.


\bibitem{capozziello}
S. Capozziello, M. De Laurentis, S.D. Odintsov, A. Stabile, \i{Phys. Rev. D} \b{83} (2011) 064004;\\
S. Capozziello, M. De Laurentis, 
\i{Phys. Rept.} {\bf 509} (2011) 167;\\
S. Capozziello, M. De Laurentis, I. De Martino, M. Formisano, S.D. Odintsov,
\i{Phys. Rev. D} {\bf 85} (2012) 044022.







\end{thebibliography}
\end{document}